\documentclass[aip,pra,preprint]{revtex4-1}

\usepackage{amsmath}
\usepackage{relsize}
\usepackage{url} 
\usepackage[colorlinks]{hyperref}
\usepackage{color}

\begin{document}

\author{Eric D'Avignon\footnote{cavell@physics.utexas.edu}}
\affiliation{Physics Department and Institute for Fusion Studies, The University of Texas at Austin, 
 Austin, TX 78712--1192}
\title{Physical Consequences of the Jacobi Identity}
\begin{abstract}
Assuming special relativity and Hamiltonian particle dynamics for a noncanonical Poisson bracket, the Jacobi identity is shown to have nontrivial physical consequences, including the homogeneous Maxwell equations and the geodesic law of motion of curved spacetime.
\end{abstract}

\maketitle

In Hamiltonian physics the Poisson bracket, whether canonical or not, must obey the Jacobi identity,
\begin{equation} \label{JacobiIdentity}
[f,[g,h]] + [g,[h,f]] + [h,[f,g]] = 0 \; .
\end{equation}
A physicist usually encounters this identity in the context of canonical brackets or quantum-mechanical operators, cases in which the proof of the Jacobi identity is a simple algebraic exercise.  However, if one approaches Hamiltonian systems from another angle, the Jacobi identity proves to have rich physical consequences.  This alternative approach begins with noncanonical, but physically meaningful, variables, in which the Jacobi identity is no longer a triviality.

Throughout this paper I will look at systems of particles, starting with a single particle.  My approach begins with a mere two assumptions: first, that the particles obey special relativity; second, that they form a Hamiltonian system.  The first assumption provides a handy set of 4-vectorial quantities describing the motion of a particle, and I will take the two simplest, spacetime position $X^\mu$ and 4-velocity $U^\mu$, as my basic dynamical variables.  This paper treats those forces that can be expressed using these variables alone, and neglects the additional forces that come from extra internal degrees of freedom like multipole moments or spinor components.

In early presentations of this work the second assumption, that the system is Hamiltonian, proved controversial.  However, past experience teaches us that a wide variety of nondissipative physical systems can be represented in Hamiltonian form, including surprising ones such as Euler's equations for a fluid, ideal MHD, and rigid body rotations \cite{morrison98}.  Moreover, quantum-mechanical systems start in a quasi-Hamiltonian form, with commutation relations replacing brackets, and the classical limits of such quantum-mechanical systems are Hamiltonian ones.  So it is a reasonable assumption: but what does it entail?  

A Hamiltonian system has two ingredients, the Hamiltonian function itself (denoted $H$), and the Poisson bracket; together, they give Hamilton's equations in their relativistic form,
\begin{equation} \label{HamiltonsEquations}
\frac{d f}{d \tau} = [f,H] \; .
\end{equation}
The $d/d\tau$ denotes the derivative with respect to the particle's proper time.  Note that this is not a $3+1$ Hamiltonian system, as in quantum field theory (see for example \cite{peskinschroeder}).  The bracket itself is fully tensorial and relies on no specific choice of reference frame, whereas a $3+1$ splitting will pull out a derivative with respect to coordinate (not proper) time by choosing a frame.  The bracket, in fact, will itself contain partial coordinate time derivatives.  

It possesses the following properties (where $f$, $g$, $h$ are functions and $\alpha$, $\beta$ are numbers), in addition to the Jacobi identity:
\begin{align*}
[\alpha f + \beta g,h] & = \alpha [f,h] + \beta [g,h] \\
[f,g] & = -[g,f] \\
[fg,h] & = f[g,h] + [f,h]g
\end{align*}
The bracket is a generalization of the partial derivative, and various properties of such derivatives, such as their linearity and Leibniz property, have analogues above.  The Jacobi identity \eqref{JacobiIdentity}, meanwhile, generalizes the commutation law of partial derivatives, 
\begin{equation*}
\frac{\partial^2 f}{\partial x_i \partial x_j} = \frac{\partial^2 f}{\partial x_j\partial x_i} \; .
\end{equation*}
Indeed, for the canonical case the Jacobi identity only requires said commutation.  In some circumstances, this commutation law can be interpreted as an integrability condition; for instance, in three dimensions, the fact that a curl-less vector field can be ``integrated'' and expressed as the divergence of a scalar relies on the commutation of partial derivatives.  Similarly, the Jacobi identity may be seen as an integrability condition.  For instance, suppose a bracket $[f,g]$ is degenerate, in that it can only generate motion along a subspace of the full variable space, irrespective of the Hamiltonian function.  This degeneracy condition may be expressed as a partial differential equation, and the Jacobi identity guarantees that this equation can be integrated to give a so-called Casimir invariant $C$, a quantity such that $[f,C] = 0$ for all functions $f$.  The forces considered in this paper will not require a degenerate bracket, but will nonetheless produce interesting integrability conditions via the Jacobi identity.

Let us begin with the simplest possible relativistic force, acting on a particle with position $X^\mu$ and velocity $U^\mu$.  Special relativity requires that the inner product $U_\mu U^\mu$ be constant; here, using normalized units, it simply equals one.  Using the 4-acceleration $K^\mu$, this condition requires
\begin{equation} \label{VelocityAccelerationOrthogonality}
U_\mu K^\mu \equiv U_\mu \frac{d U^\mu}{d \tau} = 0 \; .
\end{equation}
This orthogonality condition rules out any acceleration $K^\mu$ independent of $U^\mu$, because by varying $U^\mu$ while leaving $K^\mu$ fixed one can show the latter to be zero.  Thus, the simplest force is one linear in 4-velocity, whose archetypal example is the Lorentz force
\begin{equation} \label{LorentzForce}
K^\mu \equiv \frac{d U^\mu}{d \tau} =  \frac{e}{m} F^{\mu\nu}U_\nu \; .
\end{equation}
where $e$ is the particle's charge and $m$ is its rest mass.  The factor of $e/m$ is not strictly necessary, as I could incorporate it into the field tensor $F^{\mu\nu}$ when looking at just one particle, but I will leave it for clarity's sake.  This time the orthogonality relation \eqref{VelocityAccelerationOrthogonality} does not eliminate the force, although it does require $F^{\mu\nu}$ to be antisymmetric.  Meanwhile, the equation of motion for $X^\mu$ is simply the definition of 4-velocity:
\begin{equation} \label{Velocity}
\frac{d X^\mu}{d \tau} = U^\mu
\end{equation}

My goal is to put the equations of motion \eqref{LorentzForce} and \eqref{Velocity} into Hamiltonian form.  The quantity $U^\mu U_\mu$ is a dynamical invariant, so it must commute with the Hamiltonian, i.e. $[U_\mu U^\mu, H] = 0$.  The simplest way to assure this fact is to simply use $U_\mu U^\mu$ in the Hamiltonian, setting
\begin{equation} \label{Hamiltonian}
H = \frac{1}{2} U_\mu U^\mu
\end{equation}
In fact, any nonzero power of $U_\mu U^\mu$ can be used, changing only various numerical factors in the Hamiltonian and bracket.  I have chosen the one that produces the simplest bracket, although $H = \sqrt{U_\mu U^\mu}$ is another natural choice.  One can also add a factor of $m$ to give the Hamiltonian its usual units, but this too will complicate the bracket.  Later forces will require a more general Hamiltonian to be chosen, but for now this simple one will work.

Having chosen a Hamiltonian, I next need to construct the bracket $[f,g]$.  I will do so by finding the ``basis'' brackets $[X^\mu, X^\nu]$, $[X^\mu, U^\nu]$, and $[U^\mu, U^\nu]$, treating $X^\mu$ and $U^\mu$ as the two 4-vectors from which all other quantities can be constructed.  More general brackets can be found in either of two ways: first, if the function $f$ and $g$ are analytic, the bracket $[f,g]$ can be found by repeated use of the linearity and Leibniz properties; second, once one has the basis brackets one can infer a bracket expressed in terms of partial derivatives, as done in \eqref{ExpandedBracket} below.  I will mainly use the first approach, because then the general Jacobi identity requires only that I check four specific Jacobi identities:
\begin{equation} \begin{aligned} \label{JacobiBases}
[[X^\mu,X^\nu],X^\lambda] + [[X^\nu,X^\lambda],X^\mu] + [[X^\lambda,X^\mu],X^\nu] & = 0 \\
[[X^\mu,X^\nu],U^\lambda] + [[X^\nu,U^\lambda],X^\mu] + [[U^\lambda,X^\mu],X^\nu] & = 0 \\
[[X^\mu,U^\nu],U^\lambda] + [[U^\nu,U^\lambda],X^\mu] + [[U^\lambda,X^\mu],U^\nu] & = 0 \\
[[U^\mu,U^\nu],U^\lambda] + [[U^\nu,U^\lambda],U^\mu] + [[U^\lambda,U^\mu],U^\nu] & = 0
\end{aligned} \end{equation}

For the first piece of the bracket, I set $[X^\mu, X^\nu] = 0$.  To justify this: in differential geometry, the four vector fields $X^\mu$ spanning spacetime can be used to define a coordinate system, and one can move a finite coordinate distance along one coordinate line by exponentiating the corresponding vector field.  To form a well-behaved coordinate system, what is called a ``coordinate basis'', the fields must form a well-defined mesh: if you move by $\lambda_1$ along one vector field, then by $\lambda_2$ along a second, you should reach the same point as if you had done those two steps in the opposite order.  The necessary and sufficient condition for this desireable behavior to occur is that each commutator $[X^\mu,X^\nu]$ vanishes (see Schutz\cite{schutz} section 2.15); moreover, by Frobenius' Theorem, one can start from any set of vector fields spanning a manifold, in our case spacetime, and construct a set of commuting vector fields.  Thus, even though I am dealing with Poisson brackets instead of commutators, I will take $[X^\mu, X^\nu] = 0$ to be a defining feature of a well-behaved coordinate system.

Next up, I have the basis bracket $[X^\mu, U^\nu]$, which I can infer from the equation of motion:
\begin{equation*}
\frac{d X^\mu}{d \tau} = [X^\mu, H] = [X^\mu, U^\nu]U_\nu = U^\mu \; .
\end{equation*}
Clearly, one should set $[X^\mu, U^\nu] = g^{\mu\nu}$, where $g^{\mu\nu}$ is (for now) the Minkowski metric.  One could presumably add in a (velocity-dependent) term orthogonal to the 4-velocity $U^\mu$, but then it becomes difficult to interpret $X^\mu$ as a coordinate.  Finally, a similar equation,
\begin{equation*}
\frac{d U^\mu}{d \tau} = [U^\mu, H] = [U^\mu, U^\nu] U_\nu  = \frac{e}{m}F^{\mu\nu}U_\nu \; ,
\end{equation*}
gives the final piece of the bracket, $[U^\mu, U^\nu] = (e/m)F^{\mu\nu}$.  Technically one could still have an extra, velocity-orthogonal piece to this portion of the bracket, but I will show near the end of the paper that the Jacobi identity rules that out.  One can also write the full bracket as 
\begin{equation} \label{ExpandedBracket}
[f,g] = g^{\mu\nu}\left(\frac{\partial f}{\partial X^\mu} \frac{\partial g}{\partial U^\nu} - \frac{\partial f}{\partial X^\mu} \frac{\partial g}{\partial U^\nu} \right) + \frac{q}{m} F^{\mu\nu} \frac{\partial f}{\partial U^\mu} \frac{\partial g}{\partial U^\nu} 
\end{equation}
clearly exhibiting its canonical and noncanonical parts, and bypassing the need for analyticity in $f$ and $g$.

Now that I have a workable, indeed a simple, bracket, I need to double-check the Jacobi identity.  While the first three equations of \eqref{JacobiBases} are easily satisfied, the last is nontrivial; one finds, after lowering indices and eliminating a superfluous $(e/m)$, that
\begin{equation} \label{HomogeneousMaxwell}
F_{\mu\nu , \lambda} + F_{\nu\lambda ,\mu} + F_{\lambda\mu ,\nu} = 0
\end{equation}
These are the four homogeneous Maxwell's equations.

Half of Maxwell's equations are thus consequences of the Jacobi identity.  As noted earlier, the Jacobi identity is an integrability condition; in this case, the integrability condition for the existence of a 4-potential $A^\mu$.  Normally the existence of this potential is taken to be axiomatic, but here it arises from the fact that particles obey Hamiltonian dynamics.  Indeed, now that we have a 4-potential $A^\mu$, it is a simple matter to canonize the system: set $P^\mu = U^\mu - (e/m) A^\mu$, and you will find $[X^\mu, P^\nu] = g^{\mu\nu}$ along with $[P^\mu, P^\nu] = 0$.  In effect, I have implemented Darboux's Theorem (see for instance Crampin and Pirani \cite{crampinpirani} Sec. 6.6), which states that any finite-dimensional Poisson bracket can be decomposed into a canonical and a degenerate part, and in the process of implementing this theorem I have derived a law of physics.

One can interpret the homogeneous Maxwell equations \eqref{HomogeneousMaxwell} as implying the nonexistence of magnetic charge.  However, my exposition so far does not yet entail so much, for it only applies to a single particle.  One can imagine different types of particles responding to wholly different fields.  So, let us generalize, and consider $n$ particles responding to $m$ distinct fields.  These distinct fields will be labelled by uppercase Roman $I$ and described by the $m$ field tensors $F^{\mu\nu}_{(I)}$.  Meanwhile, each particle will be labelled by lowercase Roman $i$, and each will have its own mass $m_{(i)}$, and will respond to the fields according to $m$ distinct charges $q_{(i,I)}$.  Putting this all together, the equations of motion for these particles will be
\begin{equation*}
\frac{d X^\mu_{(i)}}{d \tau_{(i)}} = U^\mu_{(i)} \; ; \qquad 
\frac{d U^\mu_{(i)}}{d \tau_{(i)}} = \frac{1}{m_{(i)}} \sum_{I} q_{(i,I)} F^{\mu\nu}_{(I)} U^{(i)}_{\nu}
\end{equation*}
This situation is somewhat like what holds in the electroweak and strong theories, except that those require extra spinorial degrees of freedom.

Next, construct the brackets using an analogous argument to that for one particle.  The brackets factor completely, so that one has
\begin{equation*}
[X^\mu_{(i)},X^\nu_{(j)}] = 0 \; ; \quad
[X^\mu_{(i)},U^\nu_{(j)}] = g^{\mu\nu} \delta_{ij} \; ; \quad
[U^\mu_{(i)},U^\nu_{(j)}] = \frac{\delta_{ij}}{m_{(i)}} \sum_{I} q_{(i,I)} F^{\mu\nu}_{(I)} 
\end{equation*}
Checking the Jacobi identity, as before, now shows that each individual quantity $\sum_{I} q_{(i,I)} F^{\mu\nu}_{(I)}$ obeys its own version of the homogeneous Maxwell equations \eqref{HomogeneousMaxwell}.  There are thus $n$ total such equations, one for each particle, and presumably quite a bit of redundancy among them.  Suppose, for example, ``pure'' charges of some variety exist, labelled $J$.  Then, considering the Jacobi identity for one such particle, the charge factors out, and the quantity $F^{\mu\nu}_{(J)}$ by itself satisfies an equivalent of \eqref{HomogeneousMaxwell}.  Suppose, as another example, that no particular pattern among the charges exists; then, by forming an appropriate linear combination of various versions of \eqref{HomogeneousMaxwell}, one can make all but one of the charges (in this summed equation) arbitrarily close to zero, again finding that the one field $F^{\mu\nu}_{(J)}$ obeys \eqref{HomogeneousMaxwell}.  Adding multiple charges does not allow one to escape this consequence of the Jacobi identity, but instead seems to multiply such consequences.  However, perhaps one can avoid this situation by imposing some relation among the various charges $q_{(i,I)}$.  

As a specific example, I will now consider a common theory of magnetic charge (see for example Jackson \cite{jackson} Sec. 6.11), with two types of charge, electric $q_e$ and magnetic $q_m$, and two sets of fields, one, $\mathcal{F}^{\mu\nu}$, being the dual of the other, $F^{\mu\nu}$.  The Lorentz force is now
\begin{equation*}
\mathbf{F} = q_e (\mathbf{E} + \mathbf{v} \times \mathbf{B}) + q_m(\mathbf{B} - \mathbf{v} \times \mathbf{E}) \; ,
\end{equation*}
or, in tensorial notation,
\begin{equation*}
\frac{d U^\mu}{d \tau} = \frac{1}{m} \left(q_e F^{\mu\nu} U_\nu + q_m \mathcal{F}^{\mu\nu} U_{\nu}\right)
\end{equation*}
The bracket is as usual, but for $[U^\mu, U^\nu] = (1/m) (q_e F^{\mu\nu} + q_m \mathcal{F}^{\mu\nu})$, and the Jacobi identity applies to that quantity.  As explained in the previous paragraph, if I have pure electric charges, then \eqref{HomogeneousMaxwell} will apply to $F^{\mu\nu}$, and if I have pure magnetic ones, then it will apply to $\mathcal{F}^{\mu\nu}$ -- and if I have both types independently, then there is no charge whatsoever, because I no longer have any inhomogeneous Maxwell equations.  Since we know that electric charge exists, if magnetic charge also exists, it must always appear in conjunction with electric charges; moreover, it can not appear in different species in such a way that you can add them together and get zero electric, but nonzero magnetic, charge.  Some relation must exist between the electric and magnetic charges to prevent this scenario, for instance a linear relation $\alpha q_e + \beta q_m = 0$, with both $\alpha$ and $\beta$ nonzero.  However, in this case, I can define a ``total charge''
\begin{equation*}
q = \sqrt{q_e^2 + q_m^2} = q_e \sqrt{1 + \frac{\alpha^2}{\beta^2}}
\end{equation*}
and a ``total field''
\begin{equation*}
H^{\mu\nu} = \frac{1}{\sqrt{1 + \frac{\alpha^2}{\beta^2}}}\left(F^{\mu\nu} - \frac{\alpha}{\beta} \mathcal{F}^{\mu\nu}\right)
\end{equation*}
in terms of which the acceleration is
\begin{equation*}
\frac{d U^\mu}{d \tau} = q_e \left(F^{\mu\nu} - \frac{\alpha}{\beta} \mathcal{F}^{\mu\nu}\right)U_\nu
= q H^{\mu\nu} U_\nu
\end{equation*}
I have, in effect, eliminated one species of charge via a peculiar rotation, creating the plain old EM theory with $H^{\mu\nu}$ taking the role of the standard field tensor, to which, of course, the homogeneous Maxwell equations apply via the Jacobi identity.  While I don't rule out that some other, more complex relation between electric and magnetic charge might permit the latter to exist, the odds certainly do seem to be stacked against the existence of magnetic charge.

Next I will return to the single particle as a Hamiltonian system, and attempt to implement a new force.  Having dealt with forces linear in the 4-velocity, a natural next step is some kind of quadratic force, but the approach already used quickly runs into trouble.  If I take my equation of motion to be
\begin{equation*}
\frac{d U^\mu}{d \tau} = L^{\mu\nu\lambda}U_\nu U_\lambda
\end{equation*}
then the procedure that led to \eqref{HomogeneousMaxwell} now gives $[X^\mu, X^\nu] = 0$, $[X^\mu, U^\nu] = g^{\mu\nu}$, and $[U^\mu, U^\nu] = L^{\mu\nu\lambda}U_\lambda$, if I take $L^{\mu\nu\lambda}$ to be symmetric in the last two indices.  Now the third Jacobi identity in \eqref{JacobiBases} is no longer trivial; in fact, it now states
\begin{align*}
[[X^\mu,U^\nu],U^\lambda] + [[U^\nu,U^\lambda],X^\mu] + [[U^\lambda,X^\mu],U^\nu] & = \\
[g^{\mu\nu},U^\lambda] + [L^{\nu\lambda\alpha}U_{\alpha},X^\mu] + [-g^{\lambda\mu},U^\nu] & = L^{\nu\lambda\mu} = 0
\end{align*}
eliminating the quadratic force altogether.

The Hamiltonian $H$ must be changed to solve this problem.  Taking a closer look at the Hamiltonian $(1/2)U^\mu U_\mu$ = $(1/2)g_{\mu\nu} U^\mu U^\nu$, a natural alteration presents itself: why not take the metric $g^{\mu\nu}$ to no longer be constant, but instead depend on the spacetime position $X^\mu$?  This might mean that I am using a non-Minkowski coordinate system on a flat spacetime (for instance spherical or cylindrical coordinates); even better, it might imply curved spacetime, so that I am working within general, not just special, relativity.  Constructing the bracket, identical arguments again lead to $[X^\mu, X^\nu] = 0$ and $[X^\mu, U^\nu] = g^{\mu\nu}$, but now the second of these two basis brackets poses problems for the Jacobi identity.  So I will set aside the equations of motion for the moment, and instead set out to repair the bracket by choosing a $[U^\mu, U^\nu]$ that gives the correct Jacobi identity.

The first and second Jacobi identities of \eqref{JacobiBases} are still satisfied, but the third must be addressed.  Expanding it, one finds
\begin{equation} \begin{aligned} \label{ThirdJacobiWork}
[[X^\mu,U^\nu],U^\lambda] + [[U^\nu,U^\lambda],X^\mu] + [[U^\lambda,X^\mu],U^\nu] & = \\
[g^{\mu\nu},U^\lambda] + [[U^\nu,U^\lambda],X^\mu] + [[-g^{\lambda\mu},U^\nu] & = \\
g^{\mu\nu}_{\enspace \; ,\alpha}g^{\alpha\lambda} + [[U^\nu,U^\lambda],X^\mu] - g^{\lambda\mu}_{\enspace \; ,\alpha}g^{\alpha\nu} &= 0
\end{aligned} \end{equation}
One can satisfy the identity using
\begin{equation} \begin{aligned} \label{GravityUU}
[U^\mu, U^\nu] & = g^{\mu\sigma}_{\enspace\; ,\alpha}g^{\alpha\nu}U_\sigma
- g^{\nu\sigma}_{\enspace\; ,\alpha}g^{\alpha\mu} U_\sigma + \frac{q}{m}F^{\mu\nu} \\
& = g^{\mu\alpha}g^{\nu\beta}g_{\beta \sigma,\alpha}U^{\sigma}
- g^{\nu\alpha}g^{\mu\beta}g_{\beta \sigma,\alpha}U^{\sigma} + \frac{q}{m}F^{\mu\nu}
\end{aligned} \end{equation}
as the general solution.  There still remains the fourth identity in \eqref{JacobiBases}, which, after a great deal of algebraic work, evaluates to
\begin{align*}
[[U^\mu,U^\nu],U^\lambda] + [[U^\nu,U^\lambda],U^\mu] + [[U^\lambda,U^\mu],U^\nu] & = \\
\frac{q}{m} \left(g^{\lambda \alpha} F^{\mu\nu}_{\;\enspace ; \alpha} 
+ g^{\mu \alpha} F^{\nu\lambda}_{\;\enspace ; \alpha} + g^{\nu \alpha} F^{\lambda\mu}_{\;\enspace ; \alpha}\right) & \\
+ \; g^{\mu\beta} g^{\nu\gamma} g^{\lambda \delta} U^\alpha 
\left(R_{\alpha \beta\gamma\delta} + R_{\alpha \gamma\delta\beta} + R_{\alpha \delta\beta\gamma}\right) & = 0
\end{align*}
where $R_{\alpha\beta\gamma\delta}$ is the Riemann curvature tensor.  Since I can vary $X^\mu$ and $U^\mu$ separately, the two parts must each be zero.  The part linear in $U^\mu$ is satisfied due to the cyclic identity of the Riemann tensor, strange though it is that said tensor should appear; meanwhile, the part of zeroth order in $U^\mu$ are the covariant Maxwell's equations $F_{[\mu\nu ; \lambda]} = 0$ with raised indices, the extra coefficient coefficients of the covariant derivatives resulting from cross terms in the Jacobi identity.  Plainly, using covariant indices, though less natural for $X^\mu$ and $U^\mu$, would have produced simpler expressions; however, while raising and lowering indices commutes with covariant differentiation, it does \emph{not} commute with taking brackets.

So, now I have repaired the Jacobi identity, and no longer have to worry about it.  Having determined $[U^\mu,U^\nu]$, the complete bracket implies the following equation of motion:
\begin{eqnarray*}
\frac{dU^{\mu}}{d \tau} &=& \left[U^{\mu},\frac{1}{2}g_{\nu\lambda}U^{\nu}U^{\lambda}\right] \\
&=& \frac{1}{2}\left[U^{\mu},g_{\nu\lambda}\right]U^{\nu}U^{\lambda} + 
\frac{1}{2}g_{\nu\lambda}\left[U^{\mu},U^{\nu}\right]U^{\lambda} + 
\frac{1}{2}g_{\nu\lambda}\left[U^{\mu},U^{\lambda}\right]U^{\nu} \\
&=& \frac{1}{2} \left( -g^{\alpha\mu}g_{\nu\lambda,\alpha}U^{\nu}U^{\lambda} 
+ (g^{\alpha\mu}g_{\lambda\sigma,\alpha} - g^{\mu\beta}g_{\beta\sigma,\lambda})U^{\sigma}U^{\lambda} 
+ (g^{\alpha\mu}g_{\nu\sigma,\alpha} - g^{\mu\beta}g_{\beta\sigma,\nu})U^{\sigma}U^{\nu} \right) \\
&& + \; \frac{q}{m}F^{\mu\nu}U_\nu \\
&=& -\frac{1}{2}g^{\mu\alpha}\left(g_{\alpha\sigma,\lambda} + g_{\alpha\lambda,\sigma} - g_{\sigma\lambda , \alpha}
\right) U^{\sigma}U^{\lambda} + \; \frac{q}{m}F^{\mu\nu}U_\nu \\
&=& -\Gamma^{\mu}_{\;\; \sigma\lambda}U^{\sigma}U^{\lambda} + \; \frac{q}{m}F^{\mu\nu}U_\nu
\end{eqnarray*}
using the standard definition of the connection coefficients $\Gamma^{\mu}_{\enspace \sigma\lambda}$.  The Jacobi identity implies, of all things, the geodesic law of motion.  As in the EM case, one can manually implement Darboux's Theorem and make the system canonical; this time, one does so by setting $P_\mu = g_{\mu\nu}U^\nu + (e/m)A_\mu$ and finding $[X^\mu,P_\nu] = \delta^{\mu}_{\; \nu}$, $[P_\mu,P_\nu] = 0$.  Note, however, that this canonization comes at the cost of separating the phase space into contravariant $X^\mu$ and covariant $P_\mu$ parts: if you also lower $X_\mu = g_{\mu\nu} X^\nu$, the system once again becomes noncanonical.

There is another straightforward way to produce a force quadratic in 4-velocity.  Reintroduce the particle's mass $m$, which has so far been disregarded; furthermore, allow this mass to vary.  This variation might be due to internal degrees of freedom; for instance, if it has a magnetic dipole moment, expressed in an antisymmetric tensor $M^{\mu\nu}$, then the mass might be expressed as $m = m_0 + (1/2)F^{\mu\nu}M_{\mu\nu}$.  Alternatively, the system may be a composite particle where internal energy must be taken into account.  In any such situation, one introduces a nonconstant mass into the Hamiltonian as follows:
\begin{equation*}
H = \frac12 m U_{\mu}U^{\mu} - \frac12 m
\end{equation*}
where the second term ensures that $H$ still commutes with $U_{\mu}U^{\mu}$.  A factor of $(1/m)$ will now appear in the $[X^\mu,U^\nu]$ bracket, once again forcing a specific $[U^\mu,U^\nu]$ that fixes the Jacobi identity.  For simplicity, I drop any linear Lorentz-like force, and again set the metric to be constant, since these cases have already been covered, and can be reintroduced in a straightforward fashion.  The calculations needed to fix the Jacobi identity are very similar to what occurred in the prior discussion, and the resultant bracket is
\begin{align*}
\left[X^{\mu},X^{\nu}\right] &= 0 \\
\left[X^{\mu},U^{\nu}\right] &= \frac{1}{m}\eta^{\mu\nu} \\
\left[U^{\mu},U^{\nu}\right] &= \frac{1}{m^2}\left(g^{\mu\alpha}m_{,\alpha}U^{\nu} - g^{\nu\alpha}m_{, \alpha}U^{\mu}\right)
\end{align*}

This bracket yields the quadratic equation of motion 
\begin{equation*}
\frac{dU^{\mu}}{d \tau} = \left[U^{\mu},H\right] = \frac{1}{m}\left(g^{\mu\alpha}m_{,\alpha}U_{\nu}U^{\nu} - m_{,\nu}U^{\nu}U^{\mu}\right)
\end{equation*}
which is the relativistic version of the gradient force; note that, in the nonrelativistic limit, one can neglect the second term in comparison with the first, and the relation $U_\mu U^\mu = 1$ leaves nothing but a gradient.  Other quadratic forces, expressible using only $X^\mu$ and $U^\mu$, turn out to resemble combinations of the already discovered forces.  For example, throwing in a symmetric ``mass matrix'' using $H = (1/2) m_{\mu\nu} U^\mu U^\nu - (1/2)m$ turns out to have a form with a ``geodesic'' and a ``gradient'' part.

A natural next step would be to investigate further forces, cubic or higher-order in the 4-velocity, but the present endeavour comes to an end here: no matter how I modify the Hamiltonian $H$ or the basis brackets, the Jacobi identity forces any quadratic or higher term in the bracket to be zero, which in turn eliminates all cubic or higher-order forces.  Suppose, for example, I set
\begin{equation*}
[U^\mu, U^\nu] = A^{\mu\nu} + L^{\mu\nu\alpha}U_\alpha + Q^{\mu\nu\alpha\beta}U_\alpha U_\beta
\end{equation*}
Taking symmetries into account, $A^{\mu\nu}$ has 6 components, $L^{\mu\nu\alpha}$ has 24, and $Q^{\mu\nu\alpha\beta}$ 60.  However, the last Jacobi identity in \eqref{JacobiBases}, involving only velocities, will now be \emph{cubic} in $U^\mu$.  Again, the result must vanish for each order of $U^\mu$.  The zeroth order terms impose a total of 4 conditions, the linear ones 16, the quadratic ones 40, and the cubic ones 80, overdetermining the system.  This is why I ruled out terms orthogonal to $U^\mu$ in $[X^\mu, U^\nu]$ and $[U^\mu, U^\nu]$ early on, even though they would have produced the same equations of motion: these brackets would have required the use of the projection tensor $P^{\mu\nu} = g^{\mu\nu} - U^\mu U^\nu$, making the bracket at least quadratic in $U^\mu$.

The situation becomes even worse with a higher-order polynomial for $[U^\mu, U^\nu]$, for if it is $n$th order in $U^\mu$ with $n \geq 1$, the fourth Jacobi identity of \eqref{JacobiBases} will be of order $2n-1$.  One possible way around this would be to have a general Taylor series in $U^\mu$ instead of just a polynomial, because then the $n$th order of the Jacobi identity could be viewed as conditions on the $n$th term in the bracket, but I have not found a concrete example.  Of course, to introduce higher-order forces one could also add extra degrees of freedom, and thus a much more complex bracket incorporating them, but this is beyond the scope of this paper.

Nonetheless, even within the narrow scope where the arguments of this paper apply, the Jacobi identity has been shown to produce remarkable results: (i) the identity, plus a linear force, implies the homogeneous Maxwell's equations (or their equivalent); (ii) the Jacobi identity, plus a nonconstant metric, implies the geodesic law of motion; (iii) the identity, plus a nonconstant mass, produces the relativistic gradient force.  All three results are usually found by much different arguments, but bringing the Hamiltonian nature of particle motion to the fore has allowed all three to arise from one oft-neglected identity.  I hope that others will find this as remarkable as I do.

This work was supported by U.S. Dept.\ of Energy Contract \# DE-FG02-04ER54742.

\bibliographystyle{apsrev}


\begin{thebibliography}{5}
\expandafter\ifx\csname natexlab\endcsname\relax\def\natexlab#1{#1}\fi
\expandafter\ifx\csname bibnamefont\endcsname\relax
  \def\bibnamefont#1{#1}\fi
\expandafter\ifx\csname bibfnamefont\endcsname\relax
  \def\bibfnamefont#1{#1}\fi
\expandafter\ifx\csname citenamefont\endcsname\relax
  \def\citenamefont#1{#1}\fi
\expandafter\ifx\csname url\endcsname\relax
  \def\url#1{\texttt{#1}}\fi
\expandafter\ifx\csname urlprefix\endcsname\relax\def\urlprefix{URL }\fi
\providecommand{\bibinfo}[2]{#2}
\providecommand{\eprint}[2][]{\url{#2}}

\bibitem[{\citenamefont{Morrison}(1998)}]{morrison98}
\bibinfo{author}{\bibfnamefont{P.~J.} \bibnamefont{Morrison}},
  \bibinfo{journal}{Reviews of Modern Physics} \textbf{\bibinfo{volume}{70}},
  \bibinfo{pages}{467} (\bibinfo{year}{1998}).

\bibitem[{\citenamefont{Peskin and Schroeder}(1995)}]{peskinschroeder}
\bibinfo{author}{\bibfnamefont{M.~E.} \bibnamefont{Peskin}} \bibnamefont{and}
  \bibinfo{author}{\bibfnamefont{D.~V.} \bibnamefont{Schroeder}},
  \emph{\bibinfo{title}{An Introduction to Quantum Field Theory}}
  (\bibinfo{publisher}{Westview Press}, \bibinfo{year}{1995}).

\bibitem[{\citenamefont{Schutz}(1980)}]{schutz}
\bibinfo{author}{\bibfnamefont{B.}~\bibnamefont{Schutz}},
  \emph{\bibinfo{title}{Geometrical Methods of Mathematical Physics}}
  (\bibinfo{publisher}{Cambridge University Press}, \bibinfo{year}{1980}).

\bibitem[{\citenamefont{Crampin and Pirani}(1986)}]{crampinpirani}
\bibinfo{author}{\bibfnamefont{M.}~\bibnamefont{Crampin}} \bibnamefont{and}
  \bibinfo{author}{\bibfnamefont{F.}~\bibnamefont{Pirani}},
  \emph{\bibinfo{title}{Applicable Differential Geometry}}
  (\bibinfo{publisher}{Cambridge University Press}, \bibinfo{year}{1986}).

\bibitem[{\citenamefont{Jackson}(1998)}]{jackson}
\bibinfo{author}{\bibfnamefont{J.~D.} \bibnamefont{Jackson}},
  \emph{\bibinfo{title}{Classical Electrodynamics}} (\bibinfo{publisher}{John
  Wiley \& Sons, Inc.}, \bibinfo{year}{1998}).

\end{thebibliography}

\end{document}